\documentclass[letterpaper]{IEEEtran}
\usepackage{amsfonts}
\usepackage[dvips]{graphicx}
\usepackage{epsfig,latexsym}
\usepackage{float}
\usepackage{indentfirst}
\usepackage{amsmath}
\usepackage{amssymb}
\usepackage{times}
\usepackage{subfigure}
\usepackage{stfloats,algorithm,algorithmic,bm}
\usepackage{psfrag}
\usepackage{cite}
\usepackage{subfigure}

\begin{document}
\title{Ergodic Capacity Analysis of Remote Radio Head  Associations in Cloud Radio Access Networks}
\author{Mugen~Peng,~\IEEEmembership{Senior~Member,~IEEE,}
Shi~Yan, and H. Vincent Poor,~\IEEEmembership{Fellow,~IEEE}

\thanks{Manuscript received Feb. 27, 2014. The associate editor coordinating
the review of this letter and approving it for publication was Tony
Q.S. Quek.}
\thanks{Mugen~Peng (e-mail: {\tt pmg@bupt.edu.cn}), and H. Vincent Poor (e-mail:
{\tt poor@princeton.edu}) are with the School of Engineering and
Applied Science, Princeton University, Princeton, NJ, USA.
Mugen~Peng and Shi~Yan (e-mail: {\tt yanshibupt@gmail.com}) are with
the Key Laboratory of Universal Wireless Communications (Ministry of
Education), Beijing University of Posts and Telecommunications,
Beijing, China.}
\thanks{This work was supported in part by National Natural Science Foundations of China (Grant No. 61222103, No.61361166005), the National Basic Research Program of China (Grant No. 2013CB336600), the State Major Science and Technology Special Projects (Grant No. 2013ZX03001001), the Beijing Natural Science Foundation (Grant No. 4131003).}
}

\maketitle
\begin{abstract}
Characterizing user to Remote Radio Head (RRH) association
strategies in cloud radio access networks (C-RANs) is critical for
performance optimization. In this letter, the single nearest and
$N$--nearest RRH association strategies are presented, and the
corresponding impact on the ergodic capacity of C-RANs is analyzed,
where RRHs are distributed according to a stationary point process.
Closed-form expressions for the ergodic capacity of the proposed RRH
association strategies are derived. Simulation results demonstrate
that the derived uplink closed-form capacity expressions are
accurate. Furthermore, the analysis and simulation results show that
the ergodic capacity gain is not linear with either the RRH density
or the number of antenna per RRH. The ergodic capacity gain from the
RRH density is larger than that from the number of antennas per RRH,
which indicates that the association number of the RRH should not be
bigger than 4 to balance the performance gain and the implementation
cost.
\end{abstract}


\section{INTRODUCTION}
Cloud radio access networks (C-RANs) are by now recognized to
curtail both capital and operating expenditures, as well as to
provide high energy-efficiency transmission bit rates. The Remote
Radio Heads (RRHs) in C-RANs operate as soft relays by compressing
and forwarding the signals received from mobile users to a
centralized Base Band Unit (BBU) through the backhaul
links\cite{b1}. The outage probability for distributed beamforming
and best base station selection schemes in C-RANs are presented in
\cite{b2} when the user and base stations are each configured with a
single antenna and the path loss exponent is 2, and the minimal
number of RRHs for the desired user to meet a predefined quality of
service is analyzed as well. In \cite{b3}, it is demonstrated that
the large-scale fading exponent has a significant impact on the
capacity of large C-RAN systems. As an extension of \cite{b2} and
\cite{b3}, considering that the RRHs often employ multiple antennas,
the path loss exponent can vary, and the ergodic capacity is an
critical performance metric, the outage probability and ergodic
capacity performances when utilizing different RRH association
strategies for multiple antenna C-RANs are analyzed in this letter.

In particular, the aim of this letter is to study different RRH
association strategies for a user accessing the CRANs with good
reliability with constraints on implementation complexity and radio
resource consumption. The contributions are two-fold. Firstly, the
outage probability and the closed-form ergodic capacity achieved by
both the single nearest and $N$-nearest RRH association strategies
for C-RANs are characterized, where multiple antennas are used in
the RRH, and the path loss exponent is 4. Secondly, based on the
proposed outage probability and ergodic capacity performance
metrics, the impact of the number of antennas and the RRH density is
characterized. Closed-form expressions for the ergodic capacity are
derived for special cases in this letter though they are too complex
to be analyzed mathematically. According to the analysis and
simulation results, the association number of the RRH should not be
larger than 4 to balance performance gains and implementation
complexity.

\section{SYSTEM MODEL}
Consider C-RAN uplink systems, in which a group of RRHs, each having
$L$ antennas, help the signals of a single-antenna user to be
decoded in the BBU. The locations of the RRHs are assumed to be the
atoms of a two-dimensional Poisson Point Process (PPP) $\Phi$ having
intensity $\lambda$ in a disc ${\mathbb{D}}^2$, whose radius is $R$.
Without any loss of generality, we assume the desired user ( denoted
by \emph{U} ) is located at the origin of ${\mathbb{D}}^2$. Let
$\zeta(U)\in \Phi$ signify that \emph{U} is associated to an RRH.
The number $N_R$ of RRHs in ${\mathbb{D}}^2$ is random with
probability distribution $P(N_R) =
({{\mu_D}^{N_R}}/{\left({N_R}\right)!}){e^{-{\mu_D}}}$, where
${\mu_D} = \pi R^2 \lambda$. The large-scale fading is represented
by $r_i^{-\alpha}$, where $\alpha$ is the path loss exponent, and
$r_i$ is the distance between \emph{U} and the $i$-th RRH. When
maximal ratio combining (MRC) is used for achieving full-diversity
gains, the small-scale fading between \emph{U} and the \emph{i}-th
RRH is given by
\begin{equation}\label{H}
H_i = \sum\limits_{l = 1}^{{L}} {{{\left| {{h_{il}}} \right|}^2}},
\end{equation}
where $h_{il}$ is the fading gain between \emph{U} and the $l$-th
antenna of the \emph{i}-th RRH, and can be modeled as a complex
Gaussian random variable with zero mean and unit variance, i.e.
$h_{il}\sim CN(0,1)$. Thus, for the case of a large number of
antennas, $H_i$ follows the gamma distribution, i.e., $H_i\sim\Gamma
\left( {{L},1} \right)$. The probability density function (pdf) of
$H_i$ can thus be written as \linespread{0.5}
\begin{equation}\label{DH}
{f_{H_i}}\left( x \right) = \frac{{{x^{{L} - 1}}{e^{ - x}}}}{{\left(
{{L} - 1} \right)!}}.
\end{equation}
\linespread{0.5}

We let $P_U$ denote the transmit power of \emph{U}. Two RRH
association strategies are investigated in this letter.
\begin{enumerate}
  \item Single nearest RRH association: The desired user \emph{U}
  associates with the nearest RRH, which has the maximum receiving power when the shadow fading remains constant. The associated RRH $\hat{i}$ for user \emph{U} is
  thus $\hat{i} = \arg \max\limits_{i}{P_U r_i^{-\alpha} H_i}$.
  \item $N$-nearest RRH association: The desired user \emph{U}
  associates with the $N$ nearest RRHs amongst the total $N_R$ RRHs
  ($N \leq {N_R}$). To avoid the calculation of distances from \emph{U} to the total $N_R$ RRHs, the $N$ RRHs with the maximum average received power
  during the observed interval will be selected when the transmit
  powers of all RRHs are the same.
\end{enumerate}

Obviously, the higher diversity gains can be achieved by selecting
the $N$ best RRHs (i.e., the $N$ RRHs with maximum instantaneous
received power taking all kinds of fading into account) than $N$
nearest RRHs. However, to access the $N$ best RRHs, the
instantaneous channel state information (CSI) is necessary and the
backhaul signalling overhead increases with $N_R$, which challenges
the implementation complexity. Consequently, this letter focuses on
the practicable $N$-nearest RRH association strategy that selects
the $N$ RRHs with the largest received power at the BBU.

\section{Performance Analysis}

The received signal-to-noise-ratio (SNR) for \emph{U} at a distance
$r_i$ from the \emph{i}-th RRH is

\begin{equation}\label{SIR}
\gamma_{i}  = \frac{{P_U{r_i^{ - \alpha }}H_i}}{{{\sigma ^2}}},
\end{equation}
where $\sigma ^2$ is the additive thermal noise power.

\subsection{Single nearest RRH association}

For this scheme, the \emph{U} associated with the nearest RRH and
the subscript $i$ can dropped in \eqref{SIR}. An outage occurs when
the received SNR at the associated RRH is smaller than a predefined
threshold $T$.

\textbf{\emph{Lemma 1:}} \emph{The outage probability achieved by
the single nearest RRH association strategy in C-RAN uplink systems
is}
\begin{equation}\label{Outage Prob}
{\rm{P}_{out\_1R}} = \int_0^\infty \frac{\varepsilon \left(L, \frac{{{r^\alpha }T}}{\rho }\right)}{\left( L-1 \right)!}{e^{ - \lambda \pi {r^2}}}2\pi \lambda r{\rm{d}}r, \hfill\\
\end{equation}
\emph{where $\rho  = \frac{P_U}{{{\sigma ^2}}}$, and $\varepsilon
\left( {a,b} \right)$ in the numerator is the lower incomplete gamma
function given by $\varepsilon\left( {a,b} \right) = \int_0^b {{u^{a
- 1}}} {e^{ - u}}{\rm{d}}u$.}

\textbf{\emph{Proof:}} The lemma is proved in two steps: first
obtain the pdf of the distance $r$ between \emph{U} and the serving
RRH, and then find the outage probability.

Following [5], the pdf of $r$ is given by
\begin{equation}\label{Pr}
{f_r}\left( r \right) = {e^{ - \lambda \pi {r^2}}}2\pi \lambda r, r
> 0
\end{equation}

Based on \eqref{Pr}, the outage probability that the received SNR
$\gamma$ to access the nearest RRH is smaller than threshold $T$ can
be written as
\begin{equation}\label{CP}
\begin{gathered}
{\rm{P}_{out\_1R}} = \Pr \left[ {\gamma < T} \right] = {\rm{E}}\left[ {\left. {\Pr \left[ {\rho H{r^{ - \alpha }} < T} \right]} \right|r} \right]\\
\quad= \int_0^\infty \frac{\varepsilon \left(L, \frac{{{r^\alpha
}T}}{\rho }\right)}{\left( L-1 \right)!}{e^{ - \lambda
\pi{r^2}}}2\pi \lambda r{\rm{d}}r.
\end{gathered}
\end{equation}

The ergodic capacity for the proposed single nearest RRH association
strategy is specified in the following proposition:

\textbf{\emph{Proposition 1:}} \emph{For high SNR, the uplink
ergodic capacity (bps/Hz) for the single nearest RRH association
strategy in C-RAN system approximates}
\begin{equation}\label{CO3}
\begin{gathered}
 \,{C_{1R}} = \frac{{\sum\limits_{i = 1}^{L - 1} {\frac{1}{i} + \frac{\alpha }{2}\left[ {\ln \left( {\pi \lambda } \right) + C} \right]} - C  + \ln \left( P/{\sigma^2 } \right)}}{{\ln \left( 2 \right)}}, \hfill\\
   \end{gathered}
\end{equation}
\emph{where $C$ is Euler's constant.}

\textbf{\emph{Proof:}} The ergodic capacity can be calculated as
\begin{equation}\label{C}
\begin{gathered}
 {C_{1R}} =
\int_0^\infty  {{f_{\gamma_{1R}} }\left( \gamma \right){{\log }_2}(1 + \gamma){\rm{d}}\gamma},  \\
 \end{gathered}
\end{equation}
where ${f_{\gamma_{1R}} }\left(  \gamma \right)$ is the pdf of the
SNR ($\gamma$). Using the definition of pdf and the outage
probability in \eqref{CP}, we have
\begin{equation}\label{DSNR}
\begin{gathered}
{f_{\gamma_{1R}} }\left( \gamma \right) = \frac{{\partial \left( {\int_0^\infty {\Pr \left[ {H < \frac{{{r_1^\alpha }T}}{\rho }} \right]} {e^{ - \lambda \pi {r_1^2}}}2\pi \lambda r_1{\rm{d}}r_1} \right)}}{{\partial T}}.\hfill\\
 \end{gathered}
\end{equation}

Since $H\sim\Gamma \left( {{L},1} \right)$ described in \eqref{DH},
\eqref{DSNR} can be written as
\begin{equation}\label{DSNR2}
{f_{\gamma_{1R}} }\left( \gamma \right) = \int_0^\infty
{\frac{{{a^{{L}}}{{\left( \gamma \right)}^{{L} - 1}}{e^{ -
a\gamma}}}}{{\left( {{L} - 1} \right)!}}{e^{ - \lambda \pi
{r^2}}}2\pi \lambda r{\rm{d}}r},
\end{equation}
where $a=r^{\alpha}/\rho$.

In the high SNR regime, ${\log _2}\left( {1 + \gamma} \right) \sim
{\log _2}\left( \gamma \right)$. Substituting \eqref{DSNR2} into
\eqref{C}, the ergodic capacity expression can be approximated as
\begin{equation}\label{CO2}
\begin{gathered}
{C_{1R}} \approx \int_0^\infty  {\int_0^\infty  {\frac{{{a^{{L}}}{{\left( \gamma \right)}^{{L} - 1}}{e^{ - ax}}}}{{\left( {{L} - 1} \right)!}}{e^{ - \lambda \pi {r^2}}}2\pi \lambda r{\rm{d}}r{{\log }_2}(\gamma){\kern 1pt} {\rm{d\gamma}}} } \hfill \\
 \quad= \frac{1}{{\ln \left( 2 \right)}}\int_0^\infty  {\left[ {\sum\limits_{i = 1}^{L - 1} {\frac{1}{i} - C - \ln \left( {\frac{{{r^\alpha }}}{\rho }} \right)} } \right]{e^{ - \lambda \pi {r^2}}}2\pi \lambda r{\rm{d}}r} \hfill \\
\quad= \frac{{\sum\limits_{i = 1}^{L - 1} {\frac{1}{i} + \frac{\alpha }{2}\left[ {\ln \left( {\pi \lambda } \right) + C} \right]} - C  + \ln \left( P/{\sigma^2 } \right)}}{{\ln \left( 2 \right)}}. \hfill\\
\end{gathered}
\end{equation}

The derived closed-form capacity expression in \eqref{CO2} indicates
that the ergodic capacity from the single nearest RRH association
strategy is non-linearly increasing with the number of antennas per
RRH $L$, the spatially intensity of RRHs $\lambda$ and the user's
transmit power $P_U$. Furthermore, the impact on the ergodic
capacity of $\lambda$ is larger than that of $L$.

\vspace*{-1em}
\subsection{$N$-nearest RRH association}

When associating with the $N$ nearest RRHs amongst $N_R$ RRHs, the
received SNR with the MRC can be written as
\begin{equation}\label{SNR3}
\gamma_N = \sum\limits_{i = 1}^N {\frac{{PH_ir_i^{ - \alpha
}}}{{{\sigma ^2}}}}.
\end{equation}

For simplicity of description, the case $N$=2 will be presented
first, followed by the $N>2$ case.

\textbf{Case 1: Associated With 2 RRHs ($N$ = 2)}

When associating with 2 RRHs in terms of the distances of $r_1$ and
$r_2$ (assuming $0 \le r_1 \le r_2$) and the fading gains of $H_1$
and $H_2$, we have

\textbf{\emph{Lemma 2:}} \emph{The outage probability for the
2-nearest RRHs association strategy is}
\begin{equation}\label{Po32}
\begin{gathered}
{\rm{P}_{out\_2R}}= \Pr \left[ {\rho r_1^{ - \alpha }{H_1} + \rho r_2^{ - \alpha }{H_2} < T} \right]\hfill\\
\mathop  = \limits^{\left( a \right)} \int_{{{\left( {\frac{{{\rm{E}}\left\{ {\rho {H_1}} + {\rho {H_2}} \right\}}}{2}} \right)}^{\frac{1}{\alpha }}}}^\infty  {\int_{{{\left( {\frac{{{\rm{E}}\left\{ {\rho {H_1}} \right\}r_2^\alpha }}{{Tr_2^\alpha  - {\rm{E}}\left\{ {\rho {H_1}} \right\}}}} \right)}^{\frac{1}{\alpha }}}}^{{r_2}} {4{\pi ^2}{\lambda ^2}{r_1}{r_2}{e^{ - \pi \lambda r_2^2}}d{r_1}} } d{r_2} \hfill\\
\mathop  \approx \limits^{\left( b \right)} \int_{{{\left( {\frac{{2\rho {L}}}{T}} \right)}^{\frac{1}{\alpha }}}}^\infty  {2{\pi ^2}{\lambda ^2}{r_2}\left[ {r_2^2 - {{\left( {\frac{{\rho {L}r_2^\alpha }}{{Tr_2^\alpha  - \rho {L}}}} \right)}^{\frac{2}{\alpha }}}} \right]{e^{ - \pi \lambda r_2^2}}d{r_2}},  \hfill\\
\end{gathered}
\end{equation}
where $(a)$ follows from the fact that the two shortest distances
from the desired user \emph{U} are governed by the joint pdf
$f\left( {{r_1},{r_2}} \right) = 4{\pi ^2}{\lambda ^2}{r_1}{r_2}{e^{
- \pi \lambda r_2^2}}$ (\emph{Proof:} See Appendix A). Based on the
expectation of $H_i$ (i.e., $L$) in $(b)$, the double integral in
$(a)$ can be changed to be a single integral.

According to the derived single integral in \eqref{Po32}, the SNR
pdf for the 2-nearest RRH association strategry can be approximated
as
\begin{equation}\label{Pdf2}
\begin{gathered}
{f_{\gamma2R} (\gamma)} \hfill\\
\quad= \int_{{{\left( {\frac{{2{L}\rho }}{T}} \right)}^{\frac{2}{\alpha }}}}^\infty  {\frac{{2{\pi ^2}{\lambda ^2}{{\left( {{L}\rho } \right)}^{\frac{2}{\alpha }}}}}{\alpha }} {\left( {T - \left( {{L}\rho } \right){t^{ - \frac{\alpha }{2}}}} \right)^{ - \frac{2}{\alpha } - 1}}{e^{ - \pi \lambda t}}{\rm{d}}t.\hfill\\
\end{gathered}
\end{equation}

Thus, the uplink ergodic capacity can be characterized by
\eqref{C32}
\begin{figure*}[ht]
\setlength{\arraycolsep}{0.0em}
\begin{equation}\label{C32}
{C_{2R}} = \int_0^\infty  {\int_{{{\left( {\frac{{2{L}\rho }}{Z}}
\right)}^{\frac{2}{\alpha }}}}^\infty  {\frac{{2{\pi ^2}{\lambda
^2}{{\left( {{L}\rho } \right)}^{\frac{2}{\alpha }}}}}{\alpha }}
{{\left( {Z - \left( {{L}\rho } \right){t^{ - \frac{\alpha }{2}}}}
\right)}^{ - \frac{2}{\alpha } - 1}}{e^{ - \pi \lambda t}}\log
\left( {1 + Z} \right){\rm{d}}t} {\rm{d}}Z.
\end{equation}
\hrulefill
\setlength{\arraycolsep}{5pt}
\end{figure*}
located on the top of the next page. Note that this formula applies
for arbitrary $\alpha>2$, which is an extension of \cite{b2}.
Furthermore, for the case of $\alpha=4$, a simple closed-form
ergodic capacity expression can be derived as \eqref{C324},
\begin{figure*}[ht]
\setlength{\arraycolsep}{0.0em}
\begin{equation}\label{C324}
\begin{gathered}
 {C_{2R}^{\alpha=4}} =  \int_0^\infty  {\frac{{2{\pi ^2}{\lambda ^2}{{\left( {{L}\rho } \right)}^{1/2}}{e^{ - \pi \lambda t}}}}{{4\ln 2}}\left\{ {2\left[ { - \left. {\frac{{\ln Z}}{{{{ \sqrt {Z - \left( {{L}\rho } \right){t^{-2}}} }}}}} \right|_{\left( {\frac{{2{L}\rho }}{{{t^2}}}} \right)}^\infty  + \frac{2}{{{{\sqrt {\left( {{L}\rho } \right){t^{ - 2}}}}}}}\left. {\arctan \left[ {\frac{{{{\sqrt{Z - \left( {{L}\rho } \right){t^{ - 2}}}}}}}{{{{\sqrt{\left( {{L}\rho } \right){t^{ - 2}}}}}}}} \right]} \right|_{\left( {\frac{{2{L}\rho }}{{{t^2}}}} \right)}^\infty } \right]} \right\}{\rm{d}}t}  \hfill\\
 \quad \quad \;\; = \frac{{\ln \left( {2\rho {L}} \right) + \pi /2 - 2 + 2C + 2\ln \left( {\pi \lambda } \right)}}{{\ln 2}}. \hfill\\
\end{gathered}
\end{equation}
\hrulefill
\setlength{\arraycolsep}{5pt}
\end{figure*}
which shows that the number of antennas $L$ has the same impact on
the capacity as the RRH density $\lambda^2$.

\textbf{Case 2: Associated with $N$ RRHs ($N > 2$)}

To extend to the arbitrary $N$ case, the main challenge is that an
exact pdf expression for $\sum\nolimits_{i = 1}^N {r_i^{ - \alpha }}
$ is difficult to derive.  Consider the stochastic geometry property
that the points of the two-dimensional PPP of intensity $\lambda$
can be mapped to a one-dimensional PPP. The pdf of the random
variable $\pi \lambda r_i^2$ can be expressed as $f\left( x \right)
= \frac{{{x^{i - 1}}{e^{ - x}}}}{{\left( {i - 1} \right)!}}$. Hence,
the expectation of $r_i^{ - \alpha }$ can be written in
\begin{equation}\label{ER}
\begin{gathered}
{\rm{E}}\left\{ {r_i^{ - \alpha }} \right\}  = {\left( {\pi \lambda
} \right)^{\frac{\alpha }{2}}}\int_0^\infty  {{x^{ - \frac{\alpha
}{2}}}f(x){\rm{d}}x} = {\left( {\pi \lambda } \right)^{\frac{\alpha }{2}}}\frac{{\Gamma \left( {i - \frac{\alpha }{2}} \right)}}{{\Gamma \left( i \right)}}, \hfill\\
\end{gathered}
\end{equation}
where $\Gamma \left( {i - \frac{\alpha }{2}} \right)$ is finite only
for the case $i < \alpha/2$, and therefore, we should derive the
additional part when $i \ge \left\lfloor {\alpha /2} \right\rfloor+1
$, where $\left\lfloor {\cdot} \right\rfloor$ is the floor function.
The outage probability can be expressed as
\begin{equation}\label{EPoN}
\begin{gathered}
 {{\mathop{\rm P}\nolimits} _{out\_NR}} = \Pr \left[ {\sum\limits_{i = 1}^{\left\lfloor {\alpha /2} \right\rfloor } {\rho {H_i}r_i^{ - \alpha }}  + \sum\limits_{i = \left\lfloor {\alpha /2} \right\rfloor  + 1}^N {\rho {H_i}r_i^{ - \alpha }}  < T} \right] \hfill\\
 \approx \Pr \left[ {\sum\limits_{i = 1}^{\left\lfloor {\alpha /2} \right\rfloor } {\rho {H_i}r_i^{ - \alpha }}  + \rho L\sum\limits_{i = \left\lfloor {\alpha /2} \right\rfloor  + 1}^N {{{\left( {\pi \lambda } \right)}^2}} \frac{{\Gamma \left( {i - \frac{\alpha} {2}} \right)}}{{\Gamma \left( i \right)}} < T} \right]. \hfill\\
\end{gathered}
\end{equation}

In the special case of $\alpha=4$, \eqref{EPoN} can be simplified to
\begin{equation}\label{EPoN4}
\begin{gathered}
 {\mathop{\rm P}\nolimits} _{out\_NR}^{\alpha  = 4} \approx \Pr \left[ {\sum\limits_{i = 1}^2 {\rho {H_i}r_i^{ - 4}}  + \underbrace {\rho L{{\left( {\pi \lambda } \right)}^2}\frac{{N - 2}}{{N - 1}}}_S < T} \right] \hfill\\
  = \int_{{{\left( {\frac{{2\rho L}}{{T - S}}} \right)}^{\frac{1}{4}}}}^\infty {2{\pi ^2}{\lambda ^2}{r_2}\left[ {r_2^2 - \sqrt {\frac{{\rho Lr_2^4}}{{\left( {T - S} \right)r_2^4 - \rho L}}} } \right]{e^{ - \pi \lambda r_2^2}}d{r_2}}.  \hfill\\
\end{gathered}
\end{equation}

Based on \eqref{EPoN4}, the uplink ergodic capacity for the $N>2$
RRH association strategy can be derived as \eqref{Cn24}.

\begin{figure*}[ht]
\setlength{\arraycolsep}{0.0em}
\begin{equation}\label{Cn24}
\begin{gathered}
 {C_{NR}^{\alpha=4}} = \int_0^\infty  {{\pi ^2}{\lambda ^2}{e^{ - \pi \lambda t}}\sqrt {L\rho } t\left[ {\ln \left( {2L\rho  + S{t^2}} \right) - \ln \left( {{t^2}} \right) + \frac{2}{{\sqrt {L\rho  + S{t^2}} }}\arctan \left( \sqrt{L\rho  + S{t^2}} \right)} \right]{\rm{d}}t.}
\end{gathered}
\end{equation}
\hrulefill
\setlength{\arraycolsep}{5pt}
\end{figure*}

Since a closed-form pdf expression for the SNR exists in the special
case of $N \to \infty,\alpha=4$~\cite{b4} as follows
\begin{equation}\label{NSNR}
{f_\infty }\left( \gamma \right) = \frac{{\pi \lambda \sqrt {L\rho }
}}{{2{T^{3/2}}}}\exp \left( { - \frac{{L\rho {\pi ^3}{\lambda
^4}}}{{4T}}} \right),
\end{equation}
we can write an upper bound on the ergodic capacity as
\begin{equation}\label{upper}
\begin{gathered}
 {C^{Upper}} = \int_0^\infty  {\frac{{\pi \lambda \sqrt {L\rho } }}{{2{T^{3/2}}}}\exp \left( { - \frac{{L\rho {\pi ^3}{\lambda ^4}}}{{4T}}} \right){{\log }_2}\left( {1 + T} \right)} {\rm{d}}T \hfill\\
 \quad \quad \;\; \approx \frac{{C - \sum\limits_{j = 0}^\infty  {\frac{1}{{\left( {j + 1} \right)\left( {2j + 1} \right)}}}  + \ln \frac{{L\rho {\pi ^3}{\lambda ^4}}}{4}}}{{\ln 2}}. \hfill\\
\end{gathered}
\end{equation}

The derived expression in \eqref{upper} shows that the upper
capacity bound is related to the parameters $L$, $P_U$, and
$\lambda$, with $\lambda$ entering quadratically. Hence, compared
with $L$ and $P_U$, $\lambda$ is the primary factor impacting on the
ergodic capacity.

\section{NUMERICAL RESULTS}

In this section, the accuracy of the above closed-form expressions
and the impact of $\lambda$, $L$ and $P_U$ on capacity performance
are evaluated. The number of antennas per RRH $L$ is set 4, and the
expected value of $H_i$ is utilized. The path loss exponent $\alpha$
is set 4, the radius \emph{R} of the disc is set at 600m, and the
intensity of RRHs, $\lambda$, is assumed to be $10^{-4}$, i.e.,
${\mu_D} = \pi R^2 \lambda$ = 11.304. The power spectral density
$\sigma^2$ is -174dBm/Hz, and the spectral bandwidth is 100MHz.

\begin{figure}[!htp]
\centering
\includegraphics[width=4.0in]{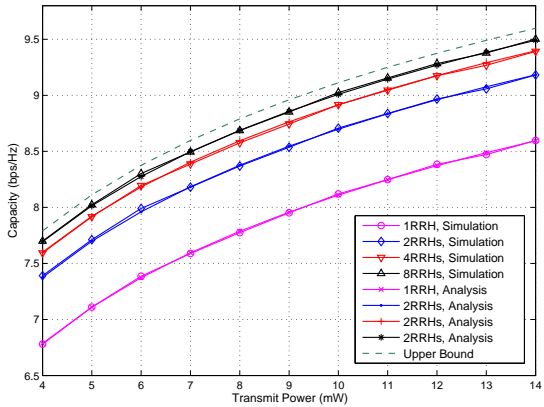}
\caption{Ergodic capacity versus transmit power
$P_U$}\label{PC}\vspace*{-1em}
\end{figure}

Fig. \ref{PC} shows the ergodic capacity performance under different
numbers of association RRHs $N$ with the varying transmit power
$P_U$, where the capacity grows monotonically as the transmit power
increases because the interference can be avoided due the
cooperative processing inherited from the C-RAN architecture. The
Monte Carlo simulation results match well with those indicated by
the presented closed-form ergodic capacity expressions. When $L = 4$
and $\lambda = 10^{-4}$, the capacity gain from the single nearest
RRH association to the 2-nearest RRH association is significant.
However, the capacity gaps among the 4 and 8 and even infinite RRH
associations are not large, which indicates that no more than 4 RRHs
should be associated for each user when considering the balance
between the performance gains and implementation cost.

\begin{figure}[!htp]
\centering
\includegraphics[width=4.0in]{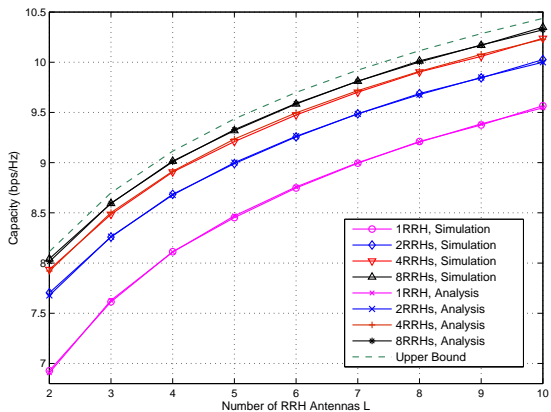}
\caption{Ergodic capacity versus antenna number $L$}\label{LC}
\vspace*{-2em}
\end{figure}

The impact of the number of antennas per RRH $L$ on the uplink C-RAN
ergodic capacity performance is shown in Fig. \ref{LC}, where $P_U =
10mw$ and $\lambda = 10^{-4}$. Similarly to the influence of the
transmission power $P_U$ shown in Fig. \ref{PC}, the uplink ergodic
capacity increases with an increasing number of antennas per RRH
$L$, and the performance gain is significant when no more than 4
RRHs are associated. Specially, when fixing $L=4$ and increasing the
RRH association number $N$ from 1 to 2 and from 4 to infinite, the
capacity performance improves about 0.58bps/Hz and 0.28bps/Hz,
respectively. The ergodic capacity for the case of $L=8$, $N=2$ is
about 9.21bps/Hz, while it is about 8.06 bps/Hz for the case of
$L=2$, $N=8$. This result demonstrates that more antennas are
preferred to increase capacity when the RRH density remains static.

\section{CONCLUSION}
In this paper, closed-form ergodic capacity expressions for both
single nearest and $N$-nearest RRH association strategies when the
pathloss exponent is 4 have been presented. Both analytical and
simulation results have shown that the RRH association number should
not be larger than 4 in order to balance the performance gain and
implementation cost, and the RRH density $\lambda$ has a greater
impact on performance gain than the number of antennas per RRH $L$
does. However, when $\lambda$ is fixed, more antennas are preferred
because this can provide higher gains than increasing RRH
association can.

\section{APPENDIX A}

We need the joint distribution probability there is not more than
one RRH within a ring from the radius $r_1$ to $r_2$, that is
\begin{equation}\label{A1}
\begin{gathered}
{\rm{Pr}}\left( {{r_1},{r_2}} \right)=\Pr \left( {null\,\in \odot {r_1},only\,one\,\in \phi_{{r_1}{r_2}}} \right) \hfill\\
\quad \quad \quad \quad \quad\cup \Pr \left( {null\,\in \odot {r_1}, null \in \phi_{{r_1}{r_2}}} \right),\hfill\\
\end{gathered}
\end{equation}
where $\odot {r_1}$ denotes the circle centered at the origin of
radius $r_1$, and $\phi_{{r_1}{r_2}}$ denotes the ring centered at
the origin of radius from $r_1$ to $r_2$. Since RRHs are distributed
according to the two-dimensional Poisson process distribution, thus
the joint probability can be written as
\begin{equation}\label{A2}
\begin{gathered}
{\rm{Pr}}\left( {{r_1},{r_2}} \right) \hfill\\
= \left( {{e^{ - \lambda \pi \left( {r_2^2 - r_1^2} \right)}} + \left( {\lambda \pi \left( {r_2^2 - r_1^2} \right)} \right){e^{ - \lambda \pi \left( {r_2^2 - r_1^2} \right)}}} \right){e^{ - \lambda \pi \left( {r_1^2} \right)}}.\hfill\\
\end{gathered}
\end{equation}


\begin{thebibliography}{99}

\bibitem{b1}
S. Park, O. Simeone, O. Sahin, and S. Shamai, ``Robust and efficient
distributed compression for cloud radio access networks,''
\emph{IEEE Transactions on Vehicular Technology}, vol. 62, no. 2,
pp. 692-703, Feb. 2013.

\bibitem{b2}
Z. Ding, and H. V. Poor, ``The use of spatially random base stations
in cloud radio access networks,'' \emph{IEEE Signal Processing
Letters}, vol. 20, no. 11, pp. 1138-1141, Nov. 2013.

\bibitem{b3}
A. Liu, and V. Lau, ``Joint power and antenna selection optimization
in large cloud radio access networks,'' \emph{IEEE Transactions on
Signal Processing}, vol. 62, no. 5, pp. 1319-1328, Mar. 2014.

\bibitem{b4}
M. Haenggi, and R. K. Ganti, ``Interference in large wireless
networks,'' \emph{Foundations and Trends in Networking}, vol. 3, no.
2, pp. 127-248, Feb. 2009.

\bibitem{b5}
J. Andrews, F. Baccelli, and R. Ganti, ``A Tractable Approach to
Coverage and Rate in Cellular Networks,'' \emph{IEEE Transactions on
Communications}, vol. 59, no. 11, pp. 3122-3134, Nov. 2011.

\end{thebibliography}
\end{document}